\documentclass[aps,twocolumn,superscriptaddress,groupedaddress,nofootinbib]{revtex4}
\usepackage{amsmath,amsfonts,amsthm,amssymb}
\usepackage[dvips]{graphics,graphicx}
\usepackage[usenames,dvipsnames]{color}
\definecolor{darkblue}{RGB}{0,0,196}
\definecolor{darkgreen}{RGB}{0,120,0}
\usepackage[colorlinks=true,linktocpage=true,linkcolor=darkblue,citecolor=red,urlcolor=darkblue]{hyperref}
\usepackage{cancel}
\usepackage{bbold}
\usepackage{multirow}
\usepackage{longtable}
\usepackage{color}
\usepackage[normalem]{ulem}
\usepackage{hyperref}
\usepackage{bigints}
\usepackage{xparse}
\usepackage{physics}
\usepackage{verbatim}
\usepackage{minibox}
\usepackage{comment}
\usepackage{appendix}
\usepackage{slashed}
\usepackage{marginnote}
\usepackage{graphicx}
\usepackage[nice]{nicefrac}
\usepackage{amsmath}
\usepackage[colorinlistoftodos]{todonotes}
\usepackage{hepunits}
\usepackage{stackengine,scalerel}
\newcommand\hstar[1]{\ThisStyle{\ensurestackMath{%
  \setbox0=\hbox{$\SavedStyle#1$}%
  \stackengine{0pt}{\copy0}{\kern.2\ht0\smash{\SavedStyle\star}}{O}{c}{F}{T}{S}}}}

\definecolor {darkgreen}{rgb}{0.2,0.7,0.2}

\begin{document}

\title{Geometric optics analysis in Lorentz violating Chern-Simons electrodynamics}

\author{Arpan Das \footnote{arpan.das@pilani.bits-pilani.ac.in}}
\address{Department of Physics,  Birla Institute of Technology and Science Pilani, Pilani Campus, Pilani,  Rajasthan-333031, India}

\author{Hriditi Howlader \footnote{p22ph005@nitm.ac.in}}
\affiliation{Department of Physics, National Institute of Technology  Meghalaya
}%

\author{Alekha C. Nayak 
\footnote{alekhanayak@nitm.ac.in}}
\affiliation{Department of Physics, National Institute of Technology Meghalaya
}%

\begin{abstract}
We study the geometric optics limit of the electrodynamics in the presence of Lorentz violating Chern-Simons term in (3+1) dimensions.   
The Chern-Simons term couples the dual electromagnetic tensor to an external four-vector and the electromagnetic gauge field. For a fixed external four-vector, such a Chern-Simons term violates Lorentz invariance while maintaining the gauge invariance of the theory. In this analysis, we look into the consequences of Lorentz symmetry violating Chern-Simons term within the geometric optics limit of light rays propagating from a source to an observation point. We argue that the Ricci tensor and the Lorentz-violating term modify the dynamical equation for the gauge vector field. However, in the geometric optics limit, neither the space-time curvature nor the Chern-Simons term influences the intensity of light. Unlike the intensity, the polarization of light, on the other hand, can be influenced by the Lorentz-violating Chern-Simons term. Due to such an effect in the presence of Chern-Simons term, the photon emitting from astrophysical objects can undergo a change in polarization as it propagates in space. 
\end{abstract}

\maketitle
\section{Introduction}
Chern-Simons term appears in a wide range of theories, e.g., topological insulators, (2+1) dimensional systems showing the Fractional Quantum Hall Effect, as well as (3+1) dimensional gauge field theories, e.g., quantum electrodynamics (QED), quantum chromodynamics (QCD), etc. In QCD the Chern-Simons term can be identified as the CP (charge conjugation and parity) violating $\theta$ term $\sim\theta G\widetilde{G}$, where $G$ denotes field strength tensor for non-abelian gauge field~\cite{Witten:1988hf,Dunne:1998qy,Bak:1994dj,TEJEROPRIETO2004138} and $\widetilde{G}$ is the dual of the field strength tensor. 
Moreover, general relativity (GR) in three dimensions can be interpreted as a topological gauge theory, and it can be described as a Chern-Simons (CS) theory~\cite{Achucarro:1986uwr,Witten:1988hc}. The Chern-Simons term in the action of the electromagnetic field can rotate polarization directions of the cosmic microwave
background (CMB)~\cite{Zhai:2020vob}. CP odd ultra-light scalars dark matter particles can couple to photon field via a Chern-Simons term analogous to the QCD $\theta$ term. Such an interaction can potentially convert the $E$-mode polarization of the CMB photon into $B$-mode polarization, resulting in new cross-correlations of the CMB polarization measurements~\cite{Fedderke:2019ajk}.  
In the context of QED, in $(3+1)$-dimensions, the Chern-Simons term can give rise to Lorentz symmetry violation without affecting the underlying gauge symmetry. This is evident because the Chern-Simons term couples the dual electromagnetic tensor ($\widetilde{F}^{\mu\nu}\equiv\epsilon^{\mu\nu\alpha\beta}F_{\alpha\beta}/2$) to an external four vector $\mathcal{P}_{\alpha}${\cite{Carroll:1989vb}}. If one considers the four-vector $\mathcal{P}_{\alpha}$ to be fixed, then it can be argued that the CS term breaks the Lorentz invariance of the theory without generating a mass term for the photon field, hence protecting the gauge symmetry~\cite{Carroll:1989vb}.

Lorentz symmetry violation has also been discussed in the context of beyond-standard model (BSM) physics. The standard model (SM) of particle physics is very successful in explaining many experimental observations. Nevertheless, it may not be a complete theory (incomplete ultraviolet theory), which motivates us to look for new theories beyond the standard model (SM) physics. One such standard model extension (SME) includes the possibility of Lorentz symmetry violation~\cite{Colladay:1998fq,Kostelecky:2003fs}. Lorentz violating theories with CPT-odd and CPT-even 
modifications related to the electromagnetic field were considered and analyzed extensively~\cite{Kostelecky:2008ts,Carroll:1989vb,Kostelecky:2001mb,Kostelecky:2002hh,Kostelecky:2006ta,Kostelecky:2009zp,Escobar:2015qla}. 
Up to the precision of current experiments, the Lorentz symmetry has been manifested as an exact symmetry. However, it has been speculated that
the deviations from Lorentz invariance could appear at an energy scale beyond the scope of the current terrestrial laboratories, e.g., at the Planck scale. Manifestation of such a Lorentz violation is reflected in the modified dispersion relation of photons~\cite{Lorentz:2016aiz}.  

In this paper, we look into the geometric optics limit of photon propagation in a general curved space-time in the presence of the Chern-Simons term in Maxwell's theory. The effect of axion field originating from such Chern-Simons term has been discussed in recent work by D. J. Schwarz et al. in Ref.~\cite{Schwarz:2020jjh}.  In the present paper, we investigate the presence of Lorentz violating Chern-Simons term in Maxwell's theory and its effects on the intensity and polarization of light rays. In this study, the metric tensor $g_{\mu\nu}$ is selected to have a mostly positive signature. In addition, we set all the fundamental constants $\epsilon_{0}$, $\mu_{0}$, $c$, and $\hbar$ to unity. The rest of the article is organized in the following manner. After this introduction in Sec.~\ref{sec2} we derive the dynamical equation of the classical electromagnetic gauge field in the presence of Lorentz violating Chern-Simons term. In Sec.~\ref{sec3}, we discuss the general framework of geometric-optics approximation, and we argue that Lorentz violating Chern-Simons term does not affect the intensity of light rays over length scales much smaller than the length scale set by the curvature characterized by the Riemann tensor. In Sec.~\ref{sec4}, we investigate the rotation of the polarization of light in the presence of Lorentz violating Chern-Simons term followed by the discussion of well celebrated Stokes parameters in Sec.~\ref{sec5}. Finally, in Sec.~\ref{sec6}, we conclude our results with an outlook to it.    

\section{Maxwell's equations with Lorentz violating Chern-Simons term}
\label{sec2}
We start with the modified action of the electromagnetic field,  
\begin{equation}
    \mathcal{S}=\int d^4x\sqrt{-g}\mathcal{L}
\end{equation}
that includes a Chem-Simons term in the Maxwell Lagrange density ($\mathcal{L}$)~\cite{Carroll:1989vb}
\begin{equation}
    \mathcal{L}=-\frac{1}{4}F_{\mu\nu}F^{\mu\nu}-\frac{g_{p}}{2}\mathcal{P}_{\alpha}A_{\beta}\widetilde{F}^{\alpha\beta}.
    \label{lagran}
\end{equation}
Here, $F_{\mu\nu}=\nabla_{\mu}A_{\nu}-\nabla_{\nu}A_{\mu}$, and $\widetilde{F}^{\alpha\beta}$ is the dual of the field Strength tensor i.e. $\widetilde{F}^{\alpha\beta}=\epsilon^{\alpha\beta\mu\nu}F_{\mu\nu}/2$. $g_p$ is a dimensionless coupling, $\nabla_{\mu}$ is the covariant derivative which is metric compatible. The external/background four-vector $\mathcal{P}_{\alpha}$ controls the Lorentz violation.  
This Lorentz violating theory is also known as the Maxwell-Carroll-Field-Jackiw (MCFJ) theory. The MCFJ model can be considered as the CPT-odd part of the extended Standard Model gauge sector~\cite{Colladay:1998fq}. 
Lorentz-violating electrodynamics has been discussed extensively in the context of topological defects solutions, quantum Hall effect, Weyl semimetals, and optical effects in a parity-violating continuous medium~\cite{Casana:2014dfa,Bailey:2004na,Gomez:2022yrp,Kostelecky:2021bsb}, etc. The above Lagrangian density gives rise to the following evolution equation of the electromagnetic gauge field ($A^{\mu}$), 
\begin{align}
& \nabla_{\mu}F^{\mu\nu}-g_p\mathcal{P}_{\mu}\widetilde{F}^{\mu\nu}=0\nonumber\\
\implies & \nabla^{\mu}\nabla_{\mu}A^{\nu}-\nabla^{\mu}\nabla^{\nu}A_{\mu}-g_p\mathcal{P}_{\mu}\widetilde{F}^{\mu\nu}=0\nonumber\\
\implies & \square A^{\nu}-\left(\nabla^{\nu}\nabla^{\mu}+R^{\mu\nu}\right)A_{\mu}-g_p\mathcal{P}_{\mu}\widetilde{F}^{\mu\nu}=0.
\label{equ3new}
\end{align}
Here $R^{\mu\nu}$ is the Ricci tensor. The Lorentz gauge choice for the gauge field $A^{\mu}$, i.e., $\nabla_{\mu}A^{\mu}=0$ allows us to simplify Eq.~\eqref{equ3new},
\begin{align}
\square A^{\nu}-R^{\mu\nu}A_{\mu}-g_p\epsilon^{\mu\nu\alpha\beta}\mathcal{P}_{\mu}\nabla_{\alpha}A_{\beta}=0. 
\label{equ3ver1}
\end{align}

\section{Geometric-optics approximation}
\label{sec3}
The geometric-optics approximation implies that the characteristic wavelength associated with gauge field configuration is much smaller than the scale of the curvature set by the Riemann tensor. Within this approximation scheme, the curvature scale is considered to be of order unity, while the wavelength of the photon field is taken to be much smaller than the curvature scale. To study the geometric optics, one proceeds with the following ansatz for the gauge field~\cite{erichill},
\begin{align}
A^{\nu} & =\left(a^{\nu}+i\epsilon b^{\nu}+\mathcal{O}(\epsilon^2)\right)e^{iS/\epsilon}.
\label{equ5new}
\end{align}
The amplitude $\left(a^{\nu}+i\epsilon b^{\nu}+\mathcal{O}(\epsilon^2)\right)$ is considered to be a slowly-varying quantity and phase factor $e^{iS/\epsilon}$ contains a rapidly-varying, real phase $S/\epsilon$. The constant parameter $\epsilon$ is taken to be small
and at the end of the one sets $\epsilon=1$, and in that case, $S$ can be identified as the actual phase. $S, a^{\alpha}$, and $b^{\alpha}$ are all space-time dependent quantities. Using the representation of the gauge field as given in Eq.~\eqref{equ5new} we find, 
\begin{align}
\nabla_{\alpha}A^{\nu} & =\left(\frac{i}{\epsilon}(\nabla_{\alpha}S )a^{\nu}-(\nabla_{\alpha}S ) b^{\nu}+\nabla_{\alpha}a^{\nu}+\mathcal{O}(\epsilon)\right)e^{iS/\epsilon}\nonumber\\
& = \left(\frac{i}{\epsilon} k_{\alpha} a^{\nu}-k_{\alpha} b^{\nu}+\nabla_{\alpha}a^{\nu}+\mathcal{O}(\epsilon)\right)e^{iS/\epsilon}.
\label{equ4ver1}
\end{align}
Here $k^{\alpha}\equiv \nabla^{\alpha}S$ can be identified as the wave vector.  Moreover the Lorentz gauge condition $\nabla_{\mu}A^{\mu}=0$ implies that, 
\begin{align}
& k_{\alpha}a^{\alpha}=0 ~~~~~~~~~\text{at}~~~\mathcal{O}(1/\epsilon),\\ 
& k_{\alpha}b^{\alpha}=\nabla_{\alpha}a^{\alpha} ~~~\text{at}~~~\mathcal{O}(\epsilon^0).
\end{align}
Therefore, only in the leading order $A^{\mu}k_{\mu}=0$, i.e., the gauge field is orthogonal to the wave vector. Using Eq.~\eqref{equ4ver1} we can proceed further to obtain, 
\begin{align}
\square A^{\nu} & =-\frac{1}{\epsilon^2}(k^{\alpha}k_{\alpha})a^{\nu}e^{iS/\epsilon}+\frac{i}{\epsilon}\bigg(-k^{\alpha}k_{\alpha}b^{\nu}+2k^{\alpha}\nabla_{\alpha}a^{\nu}\nonumber\\
& ~~~~~~~~~~~~~~~~~~~~~~~~+a^{\nu}\nabla^{\alpha}k_{\alpha}\bigg)e^{iS/\epsilon}+\mathcal{O}(\epsilon^0).
\label{equ7ver1}
\end{align}
For the geometric optics limit one considers $R^{\mu\nu}A_{\mu}=\mathcal{O}(\epsilon^0)$. Moreover Eq.~\eqref{equ4ver1} implies that, 
\begin{align}
 g_p\mathcal{P}_{\mu}\widetilde{F}^{\mu\nu}  & =g_p\epsilon^{\mu\nu\alpha\beta}\mathcal{P}_{\mu}\nabla_{\alpha}A_{\beta}\nonumber\\
& = g_p\epsilon^{\mu\nu\alpha\beta}\mathcal{P}_{\mu} \bigg(\frac{i}{\epsilon} k_{\alpha} a_{\beta}-k_{\alpha} b_{\beta}\nonumber\\
& ~~~~~~~~~~~~~~~~~~~~~~~+\nabla_{\alpha}a_{\beta}+\mathcal{O}(\epsilon)\bigg)e^{iS/\epsilon}.
\label{equ8ver1}
\end{align}
Using Eqs.~\eqref{equ7ver1}, \eqref{equ8ver1}, and $R^{\mu\nu}A_{\mu}=\mathcal{O}(\epsilon^0)$ back into Eq.~\eqref{equ3ver1}  we find that at $\mathcal{O}(1/\epsilon^2)$
\begin{align}
k^{\alpha}k_{\alpha}=0.
\label{equ9ver1}
\end{align}
Furthermore at $\mathcal{O}(1/\epsilon)$ Eq.~\eqref{equ3ver1} implies, 
\begin{align}
& 2k^{\alpha}\nabla_{\alpha}a^{\nu}+a^{\nu}\nabla_{\alpha}k^{\alpha}-g_p\epsilon^{\mu\nu\alpha\beta}\mathcal{P}_{\mu}k_{\alpha}a_{\beta}=0.
\label{equ12ver2}
\end{align}
To obtain Eq.~\eqref{equ12ver2} we explicitly use the leading order ($\mathcal{O}(1/\epsilon^2)$) condition on the wave vector, i.e., $k^{\alpha}k_{\alpha}=0$. 
The condition $k^{\alpha}k_{\alpha}=0$ also implies that, 
\begin{align}
 & k^{\alpha}\nabla_{\beta}k_{\alpha}=k^{\alpha}\nabla_{\beta}\nabla_{\alpha}S=0\nonumber\\
 & ~~~~~~~~~~~=k^{\alpha}\nabla_{\alpha}\nabla_{\beta}S=k^{\alpha}\nabla_{\alpha}k_{\beta}=0.
 \end{align}
Moreover, 
\begin{align}
\nabla_{\mu}k_{\nu}=\nabla_{\mu}\nabla_{\nu}S=\nabla_{\nu}\nabla_{\mu}S= \nabla_{\nu}k_{\mu}.
\end{align}
Contracting Eq.~\eqref{equ3ver1} with $A_{\nu}^{*}$, here  $A_{\nu}^{*}$ is the complex conjugate  $A_{\nu}^{}$, we find
\begin{align}
A^{\star}_{\nu}\square A^{\nu}-A^{\star}_{\nu}R^{\mu\nu}A_{\mu}-g_pA^{\star}_{\nu}\epsilon^{\mu\nu\alpha\beta}\mathcal{P}_{\mu}\nabla_{\alpha}A_{\beta}=0, 
\label{equ14ver1}
\end{align}
here, 
\begin{align}
A^{\star\nu} & =\left[a^{\nu}-i\epsilon b^{\nu}+\mathcal{O}(\epsilon^2)\right]e^{-iS/\epsilon}.
\label{equ16new}
\end{align}
Using Eqs.~\eqref{equ16new} and Eq.~\eqref{equ7ver1} along with the condition that $k^{\alpha}k_{\alpha}=0$ one can show that, 
\begin{align}
& A^{\star}_{\nu}\square A^{\nu}=\frac{i}{\epsilon} a^{\nu}\left(2k^{\alpha}\nabla_{\alpha}a^{\nu}+a^{\nu}\nabla^{\alpha}k_{\alpha}\right)+\mathcal{O}(\epsilon^0)
\end{align}
and, 
\begin{align}
& g_p A^{\star}_{\nu}\mathcal{P}_{\mu}\epsilon^{\mu\nu\alpha\beta}\nabla_{\alpha}A_{\beta}\nonumber\\
 = & g_p \mathcal{P}_{\mu}\epsilon^{\mu\nu\alpha\beta}\left(a_{\nu}-i\epsilon b_{\nu}+\mathcal{O}(\epsilon^2)\right)\nonumber\\
& ~~~~~~~~\times \left(\frac{i}{\epsilon}k_{\alpha}a_{\beta}-k_{\alpha}b_{\beta}+\nabla_{\alpha}a_{\beta}+\mathcal{O}(\epsilon)\right).
\end{align}
Hence at order $\mathcal{O}(1/\epsilon)$, 
\begin{align}
g_pA^{\star}_{\nu}\mathcal{P}_{\mu}\epsilon^{\mu\nu\alpha\beta}\nabla_{\alpha}A_{\beta}=0. 
\end{align}
Therefore at order $\mathcal{O}(1/\epsilon)$, Eq.~\eqref{equ14ver1} boils down to, 
\begin{align}
2 a^{\nu}k^{\alpha}\nabla_{\alpha}a^{\nu}+a_{\nu}a^{\nu}\nabla^{\alpha}k_{\alpha}=0.
\label{equ19ver1}
\end{align}
This equation can also be obtained by contacting Eq.~\eqref{equ12ver2} with the $\mathcal{O}(\epsilon^0)$ term $a^{\nu}$.  
Using the ansatz for $A^{\mu}$, and $A^{\star\mu}$ we define the intensity, 
\begin{align}
A^2=A_{\nu}^{\star}A^{\nu}=a_{\nu}a^{\nu}+\mathcal{O}(\epsilon^2),
\label{modA}
\end{align}
which allows us to write Eq.~\eqref{equ19ver1} in the following form,  
\begin{align}
k^{\mu}\nabla_{\mu}A+\frac{1}{2}A\nabla_{\mu}k^{\mu}=0.
\label{equ21ver2}
\end{align}
The above equation implies that in the geometric optics limit, neither the curvature of the space-time nor the Chern-Simons term in the Lagrangian density affects the intensity of the light source.

\section{Effect of Lorentz violating term on the polarization of light rays }
\label{sec4}
Now, let us focus on the evolution of the polarization of light rays in the presence of the Chern-Simons term. To study the polarization, we start with the following representation of the gauge field, 
\begin{align}
A^{\nu}=A\varepsilon^{\nu}e^{iS/\epsilon}.
\label{equ23new}
\end{align}
Here $\varepsilon^{\nu}$ is the normalized, space-like(complex) polarization vector i.e., $\varepsilon^{\nu}\varepsilon_{\nu}^{*}=1$ and $\varepsilon^{\nu}k_{\nu}=0$. Using Eq.~\eqref{equ23new} we find, 
\begin{align}
\square{A}^{\nu}& =\square(A\varepsilon^{\nu})e^{iS/\epsilon}+\frac{i}{\epsilon}(\nabla_{\alpha}k^{\alpha})A\varepsilon^{\nu}e^{iS/\epsilon}\nonumber\\
& ~~~~~~~~~~~+\frac{2i}{\epsilon}k_{\alpha}\nabla^{\alpha}(A\varepsilon^{\nu})e^{iS/\epsilon}.
\label{equ24new}
\end{align}
and, 
\begin{align}
& g_p\epsilon^{\mu\nu\alpha\beta}\mathcal{P}_{\mu}\nabla_{\alpha}A_{\beta}\nonumber\\
& =g_p\epsilon^{\mu\nu\alpha\beta}\mathcal{P}_{\mu}\left(\nabla_{\alpha}(A\varepsilon_{\beta})e^{iS/\epsilon}+\frac{i}{\epsilon}Ak_{\alpha}\varepsilon_{\beta}e^{iS/\epsilon}\right).
\label{equ25new}
\end{align}
Using Eqs.~\eqref{equ24new}, \eqref{equ25new} and $R^{\mu\nu}A_{\mu}=\mathcal{O}(\epsilon^0)$ back into Eq.~\eqref{equ3ver1} we find that at order $\mathcal{O}(1/\epsilon)$,
\begin{align}
& A\varepsilon^{\nu}\nabla_{\alpha}k^{\alpha}+2k_{\alpha}\varepsilon^{\nu}\nabla^{\alpha}A+2Ak_{\alpha}\nabla^{\alpha}\varepsilon^{\nu}\nonumber\\
& ~~~~~~~~~~~~~~~~~~~~-g_p\epsilon^{\mu\nu\alpha\beta}\mathcal{P}_{\mu}k_{\alpha}A\varepsilon_{\beta}=0.
\label{equ25ver2}
\end{align}
Eqs.~\eqref{equ21ver2} and \eqref{equ25ver2} allow us to write, 
\begin{align}
k^{\mu}\nabla_{\mu}\varepsilon^{\alpha}-\frac{1}{2}g_p\epsilon^{\mu\nu\rho\alpha}\mathcal{P}_{\rho}k_{\mu}\varepsilon_{\nu}=0.
\label{equ27new}
\end{align}
The above equation indicates that, unlike the intensity of the light, the polarization is affected by the presence of the Chern-Simons term. Comparing Eq.~\eqref{equ27new} with the Eq.(12) of Ref.~\cite{Schwarz:2020jjh} we observe that $\mathcal{P}_{\rho}$ plays the role of the gradient of axion field as discussed in Ref.~\cite{Schwarz:2020jjh} for the CPT-even electrodynamics. Following the framework given in Ref.~\cite{Schwarz:2020jjh} we introduce an orthogonal set of vectors $\{u,e_{1},e_{2},n\}$. 
Here $u^{\mu}$ is a time-like four-vector field with $u_{\mu}u^{\mu}=-1$, $n^{\mu}=(k^{\mu}-\omega u^{\mu})/\omega$ is a space-like four vector with  $n^{\mu}n_{\mu}=1$, $k^{\mu}u_{\mu}=-\omega$, and $e^{\nu}_{i}, i =1,2$ is a space-like linear polarisation basis. Physically 
$u^{\mu}$ describes a group of observers, $n^{\mu}$ connects the source to the group of observers, and the polarization basis $e^{\nu}_{i}, i =1,2$ covers a screen normal to $n^{\nu}$. Using the gauge freedom we can set $\varepsilon_{\nu}n^{\nu}=\varepsilon_{\nu}u^{\nu}=0$.
Then, $\varepsilon^{\nu}=\varepsilon^{i}e^{\nu}_{i}$, where $e^{i}_{\nu}e^{\nu}_{j}=\delta^{i}_{j}$ and $|\varepsilon_{1}|^2+|\varepsilon_{2}|^2=1$. Following the approach of Ref.~\cite{Schwarz:2020jjh}, using Eq.~\eqref{equ27new} we write the evolution equations of $\varepsilon^{i}$ components, 
\begin{align}
\left(\partial_{u}+\partial_{n}\right)\varepsilon^{i}-\frac{1}{2}\epsilon^{ujni}\varepsilon_{j}g_p\left(\mathcal{P}_{u}+\mathcal{P}_{n}\right)=0.
\label{equ28new}
\end{align}
To obtain the above equation one uses the condition that $k^{\mu}\nabla_{\mu}e^{\alpha}_{i}=0$. Eq.~\eqref{equ28new} is also similar to Eq.(12) of Ref.~\cite{Schwarz:2020jjh} with the identification $\partial a\rightarrow \mathcal{P}$, where $a$ is the axion field. Using the above equation, we can write the 
evolution equations for $\varepsilon^{i}$, i.e., 
\begin{align}
\varepsilon_{1}^{\prime}-\frac{1}{2}\widetilde{P}\varepsilon_{2}=0,\\
\varepsilon_{2}^{\prime}+\frac{1}{2}\widetilde{P}\varepsilon_{1}=0,
\end{align}
here the $\varepsilon_{i}^{\prime}$ denotes a derivative of $\varepsilon_{i}^{}$  along the lines of light propagation. $\widetilde{P}=g_p\epsilon^{u12n}(\mathcal{P}_{u}+\mathcal{P}_{n})$. Using $\varepsilon_{1}$ and $\varepsilon_{2}$ one can define the coefficients of
left and right circular polarisation $\varepsilon_{L,R}=(\varepsilon_{1}^{}\pm i \varepsilon_{2}^{})/\sqrt{2}$ satisfying two differential equations, 
\begin{align}
\varepsilon_{L}^{\prime}+\frac{i}{2}\widetilde{P}\varepsilon_{L}=0\\
\varepsilon_{R}^{\prime}-\frac{i}{2}\widetilde{P}\varepsilon_{R}=0.
\end{align}
The generic solution of the above set of equations can be expressed as, 
\begin{align}
\varepsilon_{L,R}^{}(x_o^{\mu}) & =\exp\left[\mp\frac{i}{2}\int_{x^{\mu}_e}^{x^{\mu}_o}\widetilde{P}~dx\right]\varepsilon_{L,R}^{}(x_e^{\mu})\nonumber\\
& = \exp(\pm \frac{i}{2}\Delta)\varepsilon_{L,R}^{}(x_e^{\mu}).
\label{lr0}
\end{align}
Here $\Delta\equiv -\int_{x^{\mu}_e}^{x^{\mu}_o}\widetilde{P}~dx$  is the rotation angle of the plane of linearly polarized light. $x^{\mu}_o$, and $x^{\mu}_e$ are the observation and emission positions of the light rays. Interestingly, in the geometric optics limit, the rotation of the polarization angle depends on the integration of the Lorentz violating term along the path of the propagation of light rays.  Also, the birefringence angle $\Delta$ is not affected by the curvature of the space-time.  This is similar to the effect of the presence of an axion field on the birefringence in the geometric optics limit, as has been shown in Ref.~\cite{Schwarz:2020jjh}. In Ref.~\cite{Schwarz:2020jjh}, it has been argued that $\Delta$ depends on the values of the axion field only at the emission and the observation point.     

\section{Stokes parameters }
\label{sec5}

In the presence of Chern-Simons term, the state of polarization may change along the direction of propagation as the photon travels from the point of emission to the point of observation.  The state of polarization can be determined by calculating the Stokes parameter.  The complete details about Stokes Parameters are given in the literature~\cite{Fedderke:2019ajk}.

Using Eq.~\eqref{lr0}, we calculate the normalized Stoke parameter $\mathcal{I}$ at the point of emission ($x^{\mu}_e$) and observation ($x^{\mu}_o$),
\begin{align}
 \mathcal{I} & = |\varepsilon_{1}|^2+|\varepsilon_{2}|^2=\varepsilon_{L}(x^{\mu}_o)\varepsilon_{L}^*(x^{\mu}_o)+\varepsilon_{R}(x^{\mu}_o)\varepsilon_{R}^*(x^{\mu}_o) \nonumber \\
 &=\varepsilon_{L}(x^{\mu}_e)\varepsilon_{L}^*(x^{\mu}_e)+\varepsilon_{R}(x^{\mu}_e)\varepsilon_{R}^*(x^{\mu}_e).
 \label{In1}
\end{align}
This implies the intensity of light does not change in the presence of Lorentz, violating Chern-Simon's terms. In the context of geometric optics, the electric field can be expressed as: $E_{\nu}=F_{\nu\beta}u^{\beta}=\mathrm{Re}[\omega A_{\nu}\exp(i \psi)]$~\cite{Schwarz:2020jjh}.  We get the intensity expression as $I=\omega^2 A^2(|\varepsilon_1|^2+|\varepsilon_2|^2)$. If we take linearly polarized light at the source of emission, then we can write $\varepsilon_2(x_e)=c ~\varepsilon_1(x_e)$, where $c$ is some real constant. Eq.~\eqref{In1} can be written as:
\begin{eqnarray}
 && \mathcal{I}_o = |\varepsilon_{1 o}|^2+|\varepsilon_{2 o}|^2=(1+c^2)|\varepsilon_1(x_e)|^2=\mathcal{I}_e
 \label{}
\end{eqnarray}
where subscript ``$o$" and ``$e$" represent the Stokes parameter at the observation and emission point, respectively.  Using Eq.\eqref{lr0}, other normalized Stokes parameters at the observation point are calculated as follows:
\begin{align}
\mathcal{Q}_o & =|\varepsilon_{1 o}|^2-|\varepsilon_{2 o}|^2\nonumber\\
& =\frac{1}{1+c^2}\left((1-c^2)\cos\Delta-2 c\sin\Delta\right)  \\
 \mathcal{U}_o & =2 \mathrm{Re}(\varepsilon_{1 o}\varepsilon_{2 o}^*)\nonumber\\
 & =\frac{1}{1+c^2}\left((1-c^2)\sin\Delta+2 c\cos\Delta\right) \\
\mathcal{V}_o & =-2 \mathrm{Im}(\varepsilon_{1 o}\varepsilon_{2 o}^*)=0 
\end{align}
The normalized Stokes parameter satisfies the relation: $\mathcal{Q}_o^{2}+\mathcal{U}_o^{2}=1$. The striking feature of the nonvanishing Lorentz-violating Chern-Simons term is that it affects the polarization due to the nonvanishing $\Delta$. In the astrophysical scenario, the light is emitted from active galactic nuclei, and the plane of polarization can change due to the Chern-Simon term. But, the intensity of light does not change. One can possibly determine the rotation angle from the observed data of distant galaxies as outlined in Ref.~\cite{Schwarz:2020jjh}. If the rotation of plane polarization is observed in the data, then we can put a constraint on the Lorentz violating parameter $\mathcal{P}_{\alpha}$ in Chern-Simon's action.

\section{Summary and Conclusions}
\label{sec6}
In this article, we explored the behavior of light in the geometric optics limit within the framework of the Maxwell-Carroll-Field-Jackiw (MCFJ) theory. The model includes a Chern-Simons term with an external four-vector, $\mathcal{P}^{\mu}$, which breaks both parity and Lorentz invariance while maintaining gauge invariance. The dynamical equation for the gauge vector field is modified by the Ricci tensor and the Lorentz-violating term. However, in the geometric optics limit, neither the space-time curvature nor the Chern-Simons term influences the intensity of light. The fact that space-time curvature has no effect reflects the nature of the geometric optics limit, where the length scale associated with the electromagnetic field is much smaller than the space-time curvature. Like intensity, the polarization of light is also not influenced by curvature (Ricci tensor); however, unlike intensity, polarization is impacted by the Lorentz-violating Chern-Simons term. To analyze the effect of this term, we also calculate the well-known Stokes parameters. Our findings show that the Lorentz-violating Chern-Simons term affects the Stokes parameters for linearly polarized light.

The presence of $\mathcal{P}^{\mu}$ in the Chern-Simons term violates the Lorentz symmetry. Taking $\mathcal{P}^{\mu} $ a constant vector, the birefringence angle ($\Delta$) depends upon the distance traveled (say $\delta  x=x_e-x_0$) by the photon. In the expanding Universe, this distance will be replaced by the comoving distance, which is given by~\cite{Jacob:2008bw,Nayak:2015xba,Weinberg:2008zzc} 
\begin{equation*}
    \delta x=\frac{1}{a_0 H_0}\int_{\frac{1}{1+z}}^1\frac{dx}{x^2\sqrt{\Omega_\Lambda+\Omega_M x^{-3}}}
\end{equation*}
where, $a_0$ is the scale factor of present epoch, $H_0$ is the Hubble constant and $z$ is the redshift of the source.  $\Omega_{\Lambda}$ is the ratio of vacuum energy density
to the critical energy density, and $\Omega_{M}$  is the ratio of non-relativistic matter density to critical energy density.
Assuming some gamma-ray bursts (GRBs) are at cosmological distances with redshift $z$, the photon emitting from such astrophysical objects undergoes a change in polarization as it propagates in space. By measuring the polarization of such a photon, the Stokes parameter can be determined. Hence, the Lorentz violating parameter can be constrained.

\section*{Acknowledgements}
 We thank Jishnu Goswami
for collaboration during the early stages of this work. A. D. acknowledges the New Faculty Seed Grant (NFSG), NFSG/PIL/2024/P3825, provided by the Birla Institute of Technology and Science Pilani, Pilani Campus, India.  A. C. N. is supported by the Science and Engineering Research Board, Republic of India/IN
under Grant No. SRG/2021/002291.

\bibliography{ref.bib}
\end{document}